\documentclass[a4paper,preprint,12pt,nofootinbib]{revtex4}
\usepackage{amsmath,mathrsfs,amscd,graphicx}
\usepackage{multirow}
\usepackage{color}
\usepackage{rotating}
\usepackage{slashed}
\usepackage{subfig}
\usepackage{textcomp}
\usepackage[final]{pdfpages}
\usepackage{array}
\usepackage{float}
\usepackage{diagbox}
\usepackage{textcomp}
\usepackage{amsfonts, latexsym, epsfig}
\usepackage{bm}
\usepackage{times}
\usepackage{epsfig}
\usepackage{amssymb}
\usepackage{tikz}
\usepackage{hyperref}
\usepackage{verbatim}
\usepackage[normalem]{ulem}
\usepackage{appendix}

\def\beq{\begin{equation}}
\def\eeq{\end{equation}}
\def\be{\begin{equation}}
\def\ee{\end{equation}}
\def\bea{\begin{eqnarray}}
\def\eea{\end{eqnarray}}
\begin{document}
\title{Probing  anomalous $WW\gamma$ Triple Gauge Bosons Coupling at the LHeC}

\author{Ruibo Li, Xiao-Min Shen, Kai Wang, Tao Xu, Liangliang Zhang and Guohuai Zhu}
\affiliation{Zhejiang Institute of Modern Physics and Department of Physics, Zhejiang University, Hangzhou, Zhejiang 310027, CHINA}

\begin{abstract}

The precision measurement of the $WW\gamma$ vertex at the future Large Hadron electron Collider (LHeC) at CERN is discussed in this paper. We propose to measure this vertex in the $e^{-} p\to e^{-}W^{\pm}j$ channel as a complement to the conventional charged current $\nu_{e}\gamma j$ channel. In addition to the cross section measurement, $\chi^{2}$ method studies of angular variables provide powerful tools to probe the anomalous structure of triple gauge boson couplings. We study the distribution of the well-known azimuthal angle between the final state forward electron and jet in this vector-boson fusion (VBF) process. On the other hand, full reconstruction of leptonic $W$ decay opens a new opportunity to measure $W$ polarization that is also sensitive to the anomalous triple gauge boson couplings. Taking into consideration the superior determination of parton distribution functions~(PDFs) based on future LHeC data, the constraints of $\lambda_{\gamma}$ and $\Delta\kappa_{\gamma}$ might reach up to $\mathcal{O}(10^{-3})$ level in the most ideal case with the 2--3~ab$^{-1}$ data set, which shows a potential advantage compared to those from LHC and LEP data. 

\end{abstract}

\maketitle

\section{Introduction}

A Standard Model~(SM) like Higgs of 125~GeV has been discovered by the ATLAS and CMS collaborations at the CERN Large Hadron Collider~(LHC)\cite{Aad:2012tfa,Chatrchyan:2012xdj}, while other hints at physics beyond the SM~(BSM) have not shown up at the current LHC run. On the other hand, there still exist many open questions that have been driving the studies of BSM physics in the last three decades. For instance, neither the mass of the Higgs boson nor the driving force of electroweak symmetry breaking~(EWSB) is explained within the SM. Therefore, precision measurements of known channels, which include precision measurement of Higgs and triple or quartic couplings of electroweak gauge boson (TGCs/QGCs) as well as rare processes of heavy flavor mesons, play an important role in the indirect probe of BSM physics.

Several electron-positron colliders such as FCC-ee, ILC and CEPC have been recently proposed as ``Higgs factories'' for the precision measurement of Higgs couplings and properties. Beside these lepton colliders, there's another relatively economic proposal for the Large Hadron electron Collider (LHeC), which is an upgrade based on the current 7~TeV proton beam of the LHC by adding one electron beam of 60--140~GeV~\cite{AbelleiraFernandez:2012cc}.  LHeC as a deep inelastic scattering (DIS) facility can improve the measurement of parton distribution at larger $x$ at TeV range significantly which is crucial for future high-energy hadron colliders.  A recent proposal of turning the machine into a Higgs factory, in which the Higgs bosons are produced via vector-boson fusion (VBF), has come out. Because of significant reduction of the QCD background and VBF forward jet tagging, the bottom quark Yukawa can be measured via $h\to b\bar{b}$~\cite{Han:2009pe}.   
In addition, by measuring via production instead of from decay, the LHeC has apparent advantages in studying anomalous $VVh$ coupling.

At the same time, there also exist several studies on anomalous TGC{s}~(aTGC{s}) at these Higgs factories.  At the LHeC, the TGC{s} can be directly probed via single $\gamma/Z$ and single $W$ production~\cite{Biswal:2014oaa,cakir,cakir1,Baur:1989gh}. In this work, we focus on the $e^{-}p\rightarrow e^{-}W^{\pm}j$ process because in this channel leptonic $W$ decay could provide its polarization information as an additional handle. That is to say, one can further use $\cos\theta_{\ell W}$, which is defined with the moving direction of decay product $\ell$ and the $W$ boson itself, to distinguish contributions from anomalous couplings. This serves as a useful complementary channel to the aTGC study in single $\gamma/Z$ production measurement, in which the total cross section and azimuthal angle distribution are usually used. 
  
In principle, the $e^{-}p\rightarrow e^-W^{\pm} j$ process contains both diagrams with the $WWZ$ vertex and diagrams with the $WW\gamma$ vertex, which interfere with each other.~However, due to large suppression from $Z$ boson mass, the results are actually insensitive to anomalous $WWZ$ couplings~\cite{Baur:1989gh}. Therefore,~{we set the anomalous $WWZ$ couplings to zero} and use the results in this study as a direct constraint on the anomalous $WW\gamma$ vertex.

This paper is organized as follows. In the next section, we discuss the physics argument of proposed differential distributions and the current status of aTGC measurement. In section~\ref{pheno} we discuss the phenomenology of this collider search, which includes event selection and reconstruction, $W$ polarization analysis and azimuthal angle correlation analysis. In section~\ref{result}, we give the numerical analysis results with the $\chi^2$ method. In the last section, we give a brief conclusion.

\section{aTGC and $W$ polarizations}

As stated in the introduction, we focus on 
$e^{-}p\rightarrow e^{-}W^{\pm}j$ 
which provides additional information on $W$-polarization as a handle besides the known azimuthal angle dependence $\Delta \phi_{ej}$. To measure the $W$ polarization, we choose the muonic decay subchannel,
\beq
e^{-} +p\to e^{-} + j +W^{\pm} \to e^{-}+j+\mu^{\pm}+\nu_{\mu}~.
\eeq
We neglect the electronic and hadronic decay channel to avoid combinatory backgrounds and additional irreducible backgrounds. Detailed discussion on this can be found in section~\ref{pheno}.

\begin{figure}[H]
\centering
\subfloat[]{\includegraphics[scale=0.33]{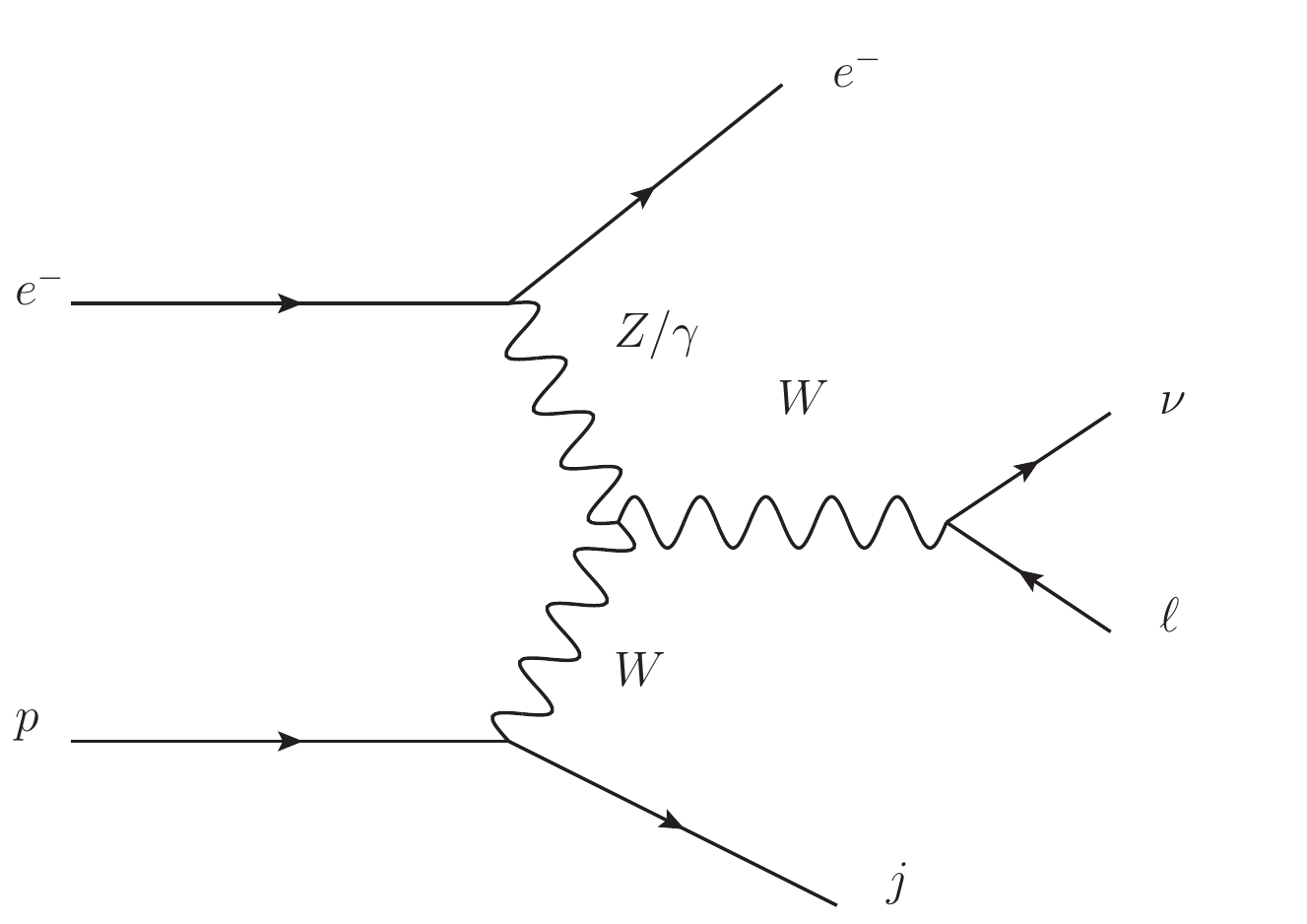}}\qquad
\subfloat[]{\includegraphics[scale=0.4]{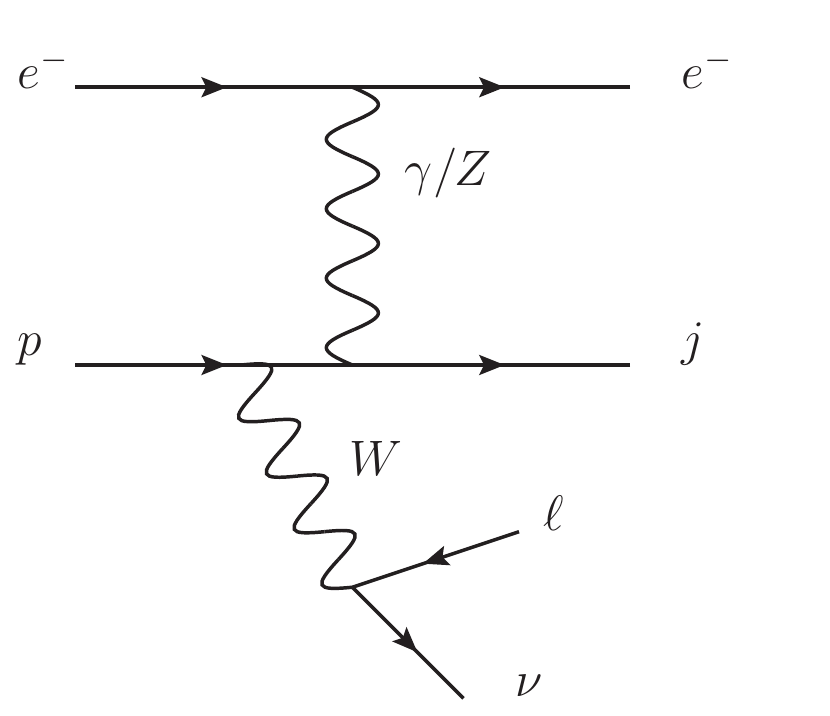}}\qquad
\subfloat[]{\includegraphics[scale=0.4]{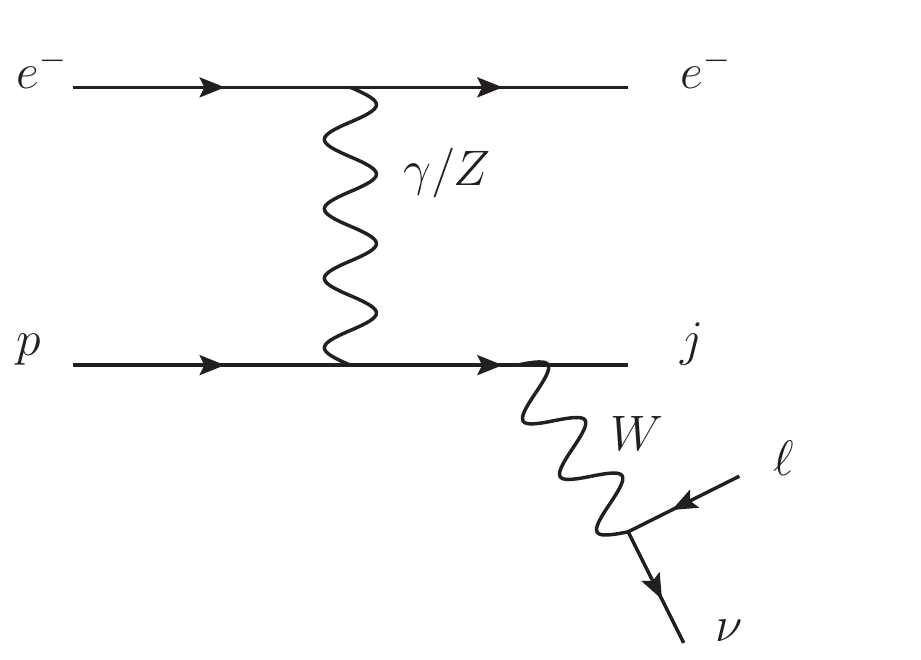}}\\
\subfloat[]{\includegraphics[scale=0.4]{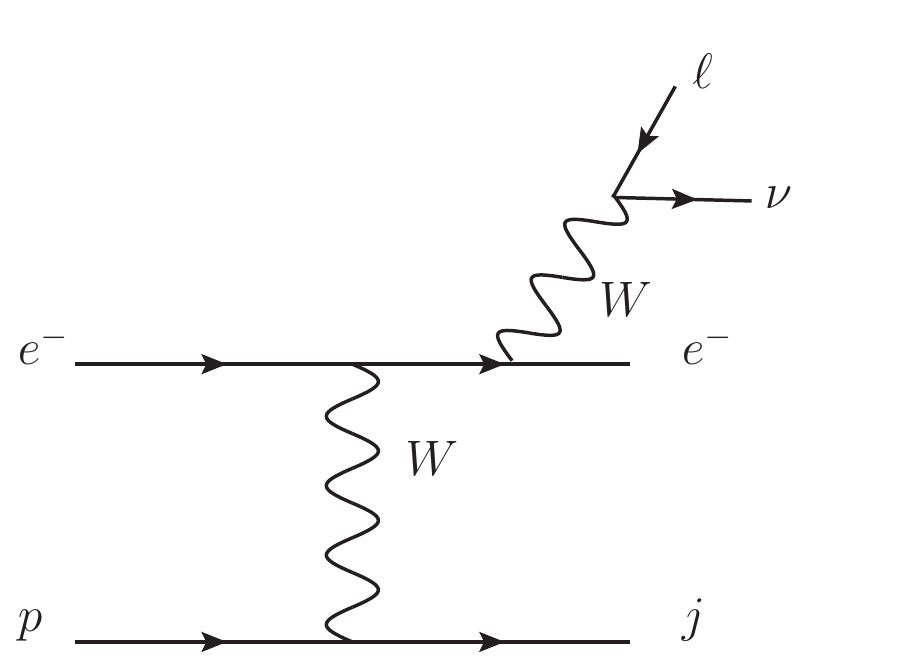}}\qquad
\subfloat[]{\includegraphics[scale=0.35]{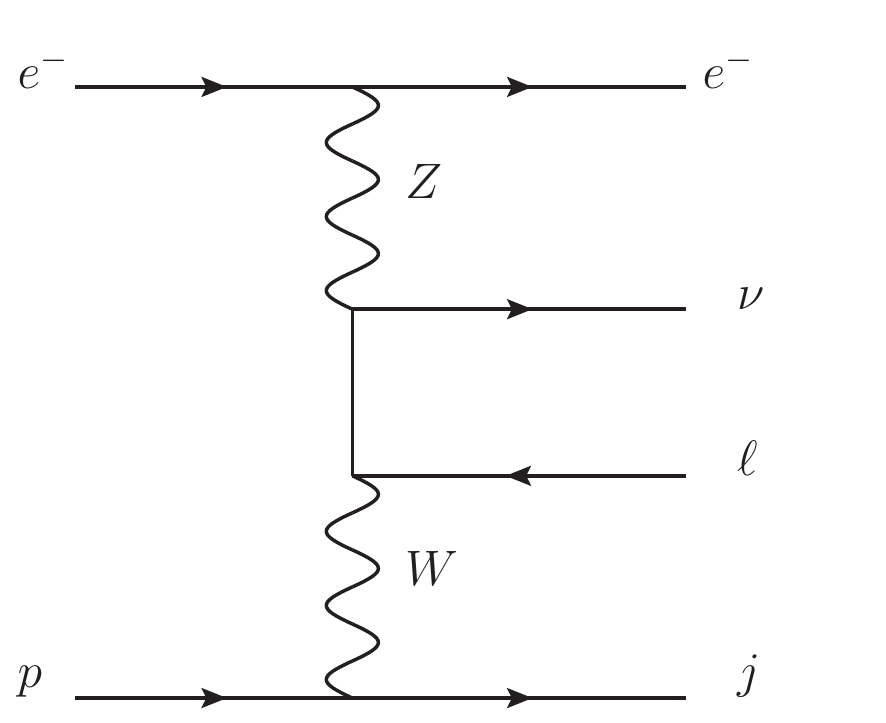}}\qquad
\subfloat[]{\includegraphics[scale=0.35]{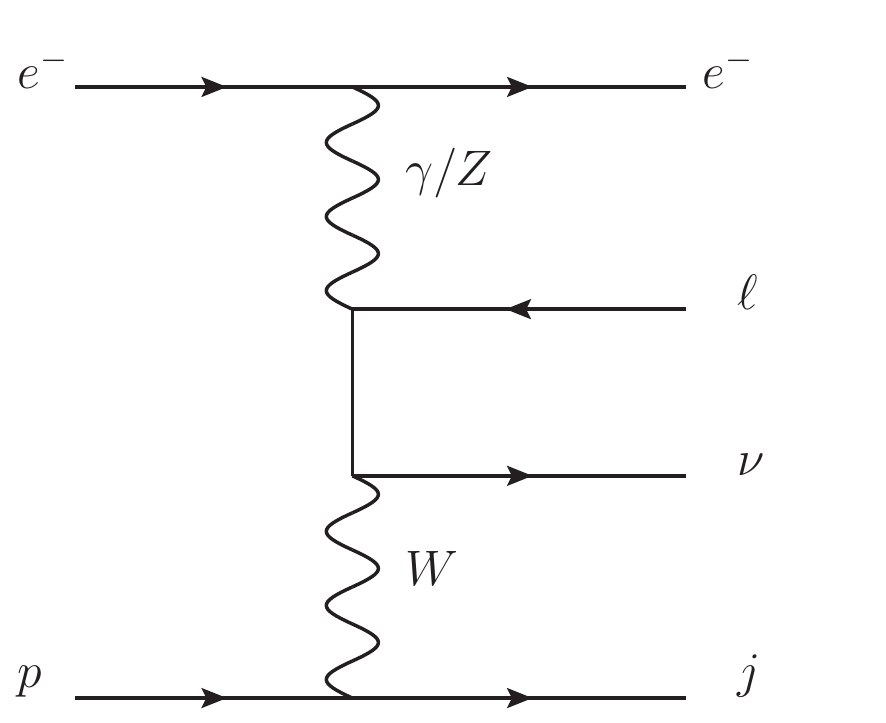}}
\caption{Diagrams of the $e^{-}p\rightarrow e^{-}\mu^{+}\nu_{\mu}j$ process. (a) is TGC contribution and (b)--(f) are backgrounds.}
\label{signals-TGCs}
\end{figure}
The diagrams contributing to $e^{-}p\rightarrow e^{-}\mu^{+}\nu_{\mu}j$ are shown in Fig.~\ref{signals-TGCs}. If one computes the single TGC-only diagram as in Fig.~\ref{signals-TGCs} (a), the longitudinal polarized $W$ dominates and the cross section is huge.~On the other hand, it is well known that the gauge symmetry unitarizes the scattering amplitudes which ensures the consistency condition of theories for spin-1 vector boson. 
In a theory with exact gauge symmetry such as QED, the requirement that current associated with the gauge symmetry is covariantly conserved leads to the result that the longitudinal-polarized component of massless vector-boson cancels and does not contribute to physical processes as Ward-Takahashi Identity, 
\beq
\partial_{\mu} { J}^{\mu} = 0 \to q_{\mu} \left<J^{\mu}(q)\right>=0~.
\eeq
This vanishing of contribution from longitudinal-polarized component also occurs when the gauge symmetry is spontaneously broken but the contribution of longitudinal polarized component is not exactly zero. 
Therefore, when all the diagrams in Figs.\ref{signals-TGCs}(b)--\ref{signals-TGCs}(f) are included, the gauge invariance is restored by a large cancellation between the longitudinal components. Such cancellation among the complete gauge-invariant set of diagrams could reduce the cross section by two orders of magnitude.  
\begin{figure}[H]
\centering
\includegraphics[width=0.45\textwidth]{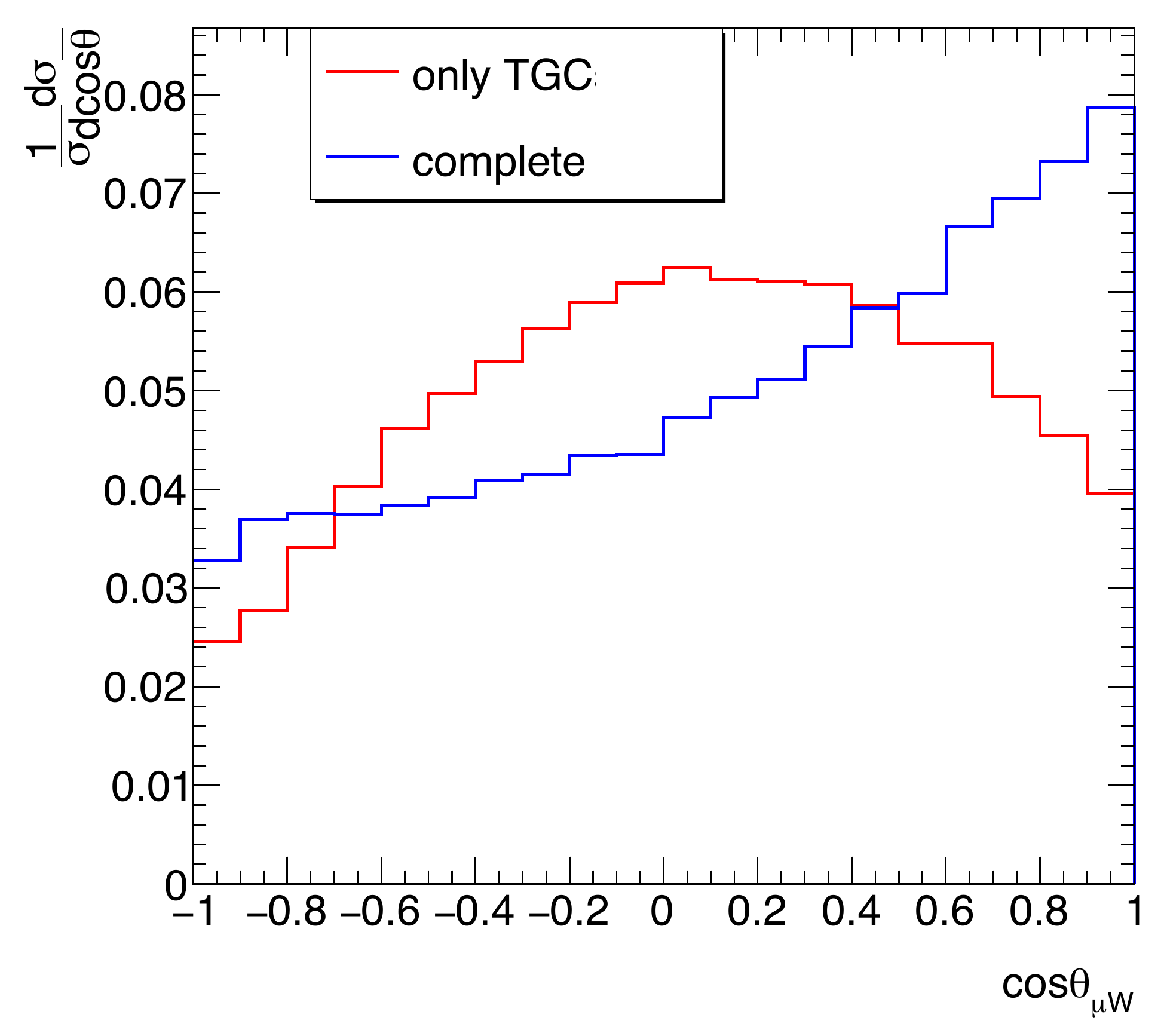}
\includegraphics[width=0.45\textwidth]{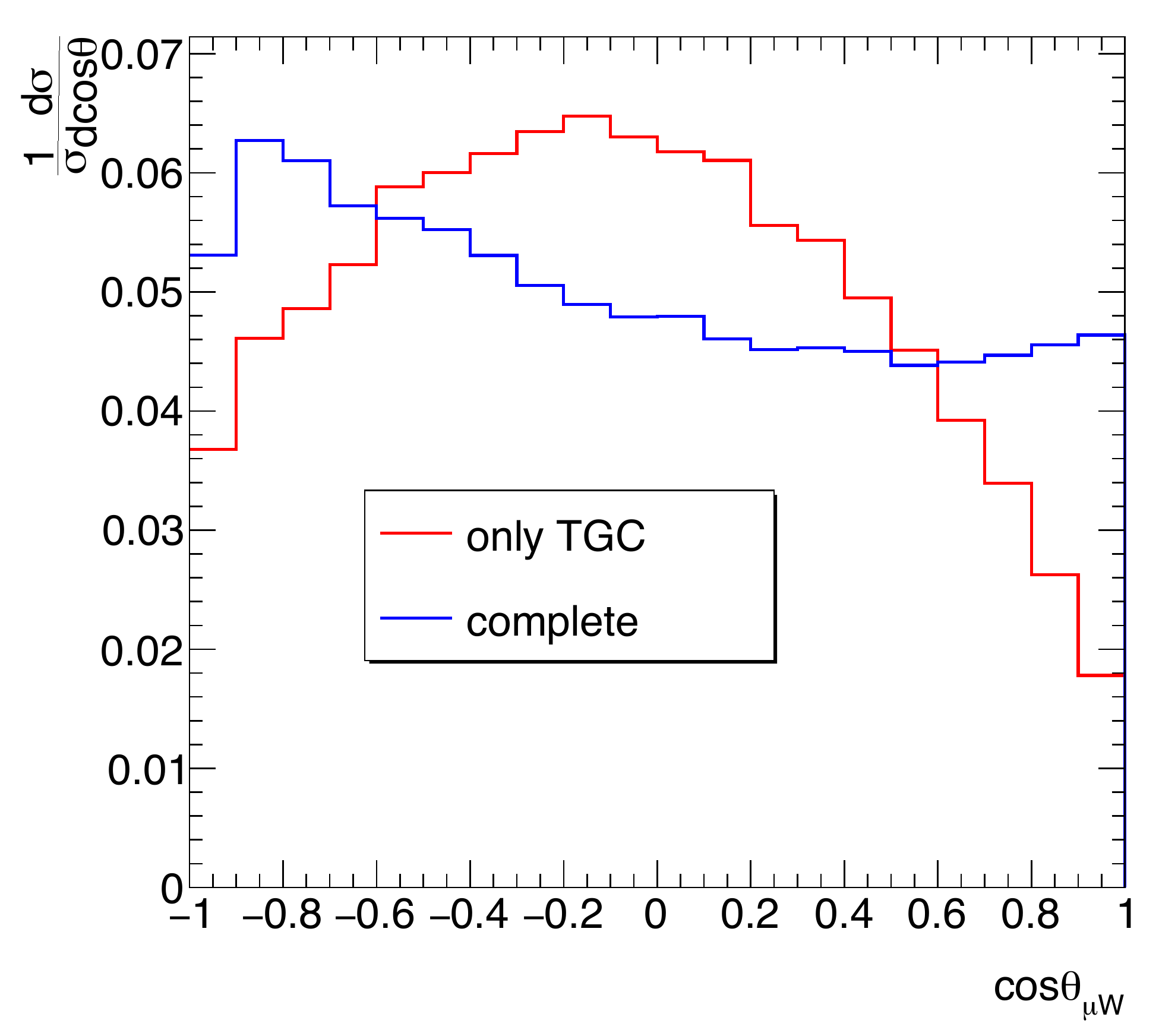}
\caption{Comparison of normalized $\cos\theta_{\mu W}$ differential distributions of the TGC graph contribution~(red) and complete contribution~(blue). The Left panel is for $e^{-}p\rightarrow e^{-}\mu^{-}\bar{\nu}_{\mu}j$ and the right panel is for $e^{-}p\rightarrow e^{-}\mu^{+}\nu_{\mu}j$. }
\label{signals-TGCs-costheta}
\end{figure}
Fig.~\ref{signals-TGCs-costheta} shows a comparison between $\cos\theta_{\mu W}$ distributions of TGC contributions and the complete set of diagrams which confirms the above argument. Hence, this differential cross section can provide additional information on the process.

The triple gauge boson vertices with anomalous contributions could be generally parametrized by effective Lagrangian as~\cite{Baur:1989gh,Bian:2015zha}
\bea
\mathcal{L}_{TGC}/g_{WWV}&=&ig_{1,V}(W_{\mu\nu}^{+}W_{\mu}^{-}V_{\nu}-W_{\mu\nu}^{-}W_{\mu}^{+}V_{\nu})+i\kappa_{V}W_{\mu}^{+}W_{\nu}^{-}V_{\mu\nu}+\frac{i\lambda_{V}}{M_{W}^{2}}W_{\mu\nu}^{+}W_{\nu\rho}^{-}V_{\rho\mu}\nonumber\\
&&+g_{5}^{V}\epsilon_{\mu\nu\rho\sigma}(W^{+}_{\mu}\overleftrightarrow{\partial}_{\rho}W^{-}_{\nu})V_{\sigma}-g_{4}^{V}W^{+}_{\mu}W^{-}_{\nu}(\partial_{\mu}V_{\nu}+\partial_{\nu}V_{\mu})\nonumber\\
&&+i\tilde{\kappa}_{V}W^{+}_{\mu}W^{-}_{\nu}\tilde{V}_{\mu\nu}+\frac{i\tilde{\lambda}_{V}}{M_{W}^{2}}W^{+}_{\lambda\mu}W^{-}_{\mu\nu}\tilde{V}_{\nu\lambda}
\label{lan-TGCs}
\eea
$V=\gamma, Z$. The gauge couplings are $g_{WW\gamma}=-e$, $g_{WWZ}=-e\cot\theta_{W}$. $\theta_{W}$ is the weak mixing angle. $\tilde{V}_{\mu\nu}$ and $A\overleftrightarrow{\partial}_{\mu}B$ are defined as $\tilde{V}_{\mu\nu}=\frac{1}{2}\epsilon_{\mu\nu\rho\sigma}V_{\rho\sigma}$, $A\overleftrightarrow{\partial}_{\mu}B=A(\partial_{\mu}B)-(\partial_{\mu}A)B$ respectively. The charge (C) and parity (P) conjugate properties of the terms in Eq.~(\ref{lan-TGCs}) are as follows. $g_{4}^{V}$ violates $C$ and $CP$, $g_{5}^{V}$ violates $C$ and $P$ but preserves $CP$, and $\tilde{\kappa}_{V}$ and $\tilde{\lambda}_{V}$ are $P$ and $CP$ violating.   
The rest of the couplings $g_{1,V}$, $\kappa_{V}$, and $\lambda_{V}$ are both $C$ and $P$ conserving. There're only five $C$ and $P$ conserving aTGCs because electromagnetic gauge symmetry requires $g_{1,\gamma}=1$. We can reduce two of them for independency because of the relations $\lambda_{\gamma}=\lambda_{Z}$ and $\Delta\kappa_{Z}=\Delta g_{1,Z}-\tan^{2}\theta_{W}\Delta\kappa_{\gamma}$~\cite{Hagiwara:1993ck, Hagiwara:1992eh, DeRujula:1991ufe}. So the only independent aTGCs are $\Delta g_{1,Z},\Delta\kappa_{\gamma}$ and $\lambda_{\gamma}$, which should vanish in the SM.

These constant aTGCs, in contrast to the SM, lead to rapid growth in the scattering cross section with collision energy until some high-energy scale $\Lambda$, where unitarity breaks down. Therefore, unitarity sets an upper bound on aTGC values for it doesn't break down before $\sqrt{s}\sim\Lambda$, and this $\Lambda$ is equivalent to the lower bound above which new physics could saturate unitarity. On the other hand, a severe aTGC constraint ensures that the effective field theory description in new physics searches is valid, throughout the energy scale our present collider experiments could reach. The bounds on aTGCs from $ff'\to VV'$ scattering unitarity are $|\Delta\kappa_{\gamma}|\leq1.86/\Lambda^2$ and $|\lambda_{\gamma}|\leq0.99/\Lambda^2$ where $\Lambda$ is in TeV~\cite{Baur:1987mt}. The cutoff scale $\Lambda$ is larger than $3$~TeV for aTGC sensitivity better than $\mathcal{O}(0.1)$. The $VV'\to VV'$ scattering also sets unitarity breaking scales from the present aTGC bound, but they're all in the several TeV range~\cite{Corbett:2017qgl}. Therefore, LHeC collision energy is safe from violating scattering unitarity and its high sensitivity to aTGC would improve the unitarity bound for future energy frontier experiments.

In Table~\ref{bounds-EXP}, we list the current  95\% C.L. bounds on aTGCs, based on diboson production measurements at LEP and LHC. At present, LHC measurements of $WW/WZ$ pair production in their semi-leptonic decay channel give the most stringent bounds~\cite{Schael:2013ita,cms-2017-tgcs, Aaboud:2017cgf}.

\begin{table}[H]
\begin{center}
\begin{tabular}{|c|c|c|c|c|}
\hline
aTGC&LEP~\cite{Schael:2013ita}&CMS, 8~TeV~\cite{cms-2017-tgcs}&ATLAS, 8~TeV~\cite{Aaboud:2017cgf}& SM\\\hline\hline
$\Delta g_{Z}$&[-0.054, 0.021]&[-0.0087, 0.024]&[-0.021, 0.024]&0\\\hline
$\Delta\kappa_{\gamma}$&[-0.099, 0.066]&[-0.044, 0.063]&[-0.061, 0.064]&0\\\hline
$\lambda_{\gamma}$&[-0.059, 0.017]& [-0.011, 0.011]&[-0.013, 0.013]&0\\\hline
\end{tabular}
\caption{95\% C.L. limits on $\Delta g_{Z}$, $\Delta\kappa_{\gamma}$ and $\lambda_{\gamma}$ at the LEP and LHC.  These bounds are from single-parameter fittings.}
\label{bounds-EXP}
\end{center}
\end{table}

\section{Phenomenology of aTGC measurements at LHeC}
\label{pheno}
\subsection{Event selection and signal production}
In this section, we discuss the collider phenomenology of aTGC measurement through the $e^{-}p\rightarrow e^{-}W^{\pm}j\rightarrow e^{-}\ell^{\pm}\nu_{\ell}j$ process and use {\it MadGraph5\_v2.4.2}~\cite{Alwall:2014hca} for a parton-level analysis of the measurements. There are four different leptonic channels. For $\ell=e^{+}$, the $e^+e^-$ pair from processes with neutral boson decay would be additional backgrounds that we want to avoid.~For $\ell=e^{-}$, the mistagging rate between the electron from $W$ boson decay and the scattered beam electron is 7\%, if we assume the electron from $W$ decay takes the smaller rapidity value. On the other hand, neutral current deep inelastic scattering events in the $e^{-}$ channel are potential sources of backgrounds as well. For $\ell=\mu^{-}$, its signal production rate would be smaller than in the $\mu^{+}$ channel because of the parton distribution of proton~($uud$) at the $e^-p$ collider. Thus among all the leptonic channels, we expect the $\mu^{+}$ channel to be more sensitive to aTGCs than others.~{With respect to $W$ hadronic decay channel,~we need to consider $e^{-}$ + 3$j$ with a $30.53~pb$ production cross section as the final state, which is approximately two orders over the leptonic decay channel because of huge QCD processes. When $E_{e}=60~\rm{GeV}$, we checked the dijet from $W$ decay would not appear as a single fat jet.~One can set the dijet-invariant mass cut and forward jet tagging as a means to reduce QCD backgrounds and extract electroweak processes, but the cross section is still $\mathcal{O}(pb)$ level despairing to probe tiny aTGC contributions. Moreover, because of the jet substructure, we can’t define the polar angle between decay product jets and W boson. Therefore, the W boson polarization information we focus on could no longer be used. }

\begin{figure}[H]
\centering
\subfloat[]{\includegraphics[scale=0.6]{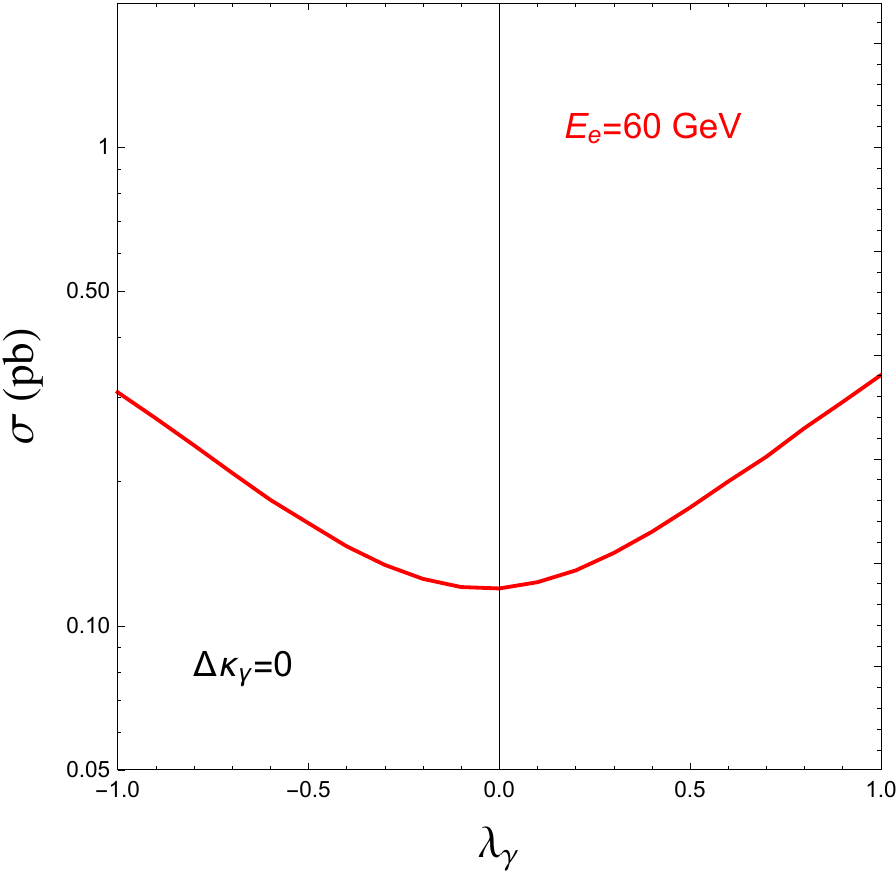}}\qquad
\subfloat[]{\includegraphics[scale=0.6]{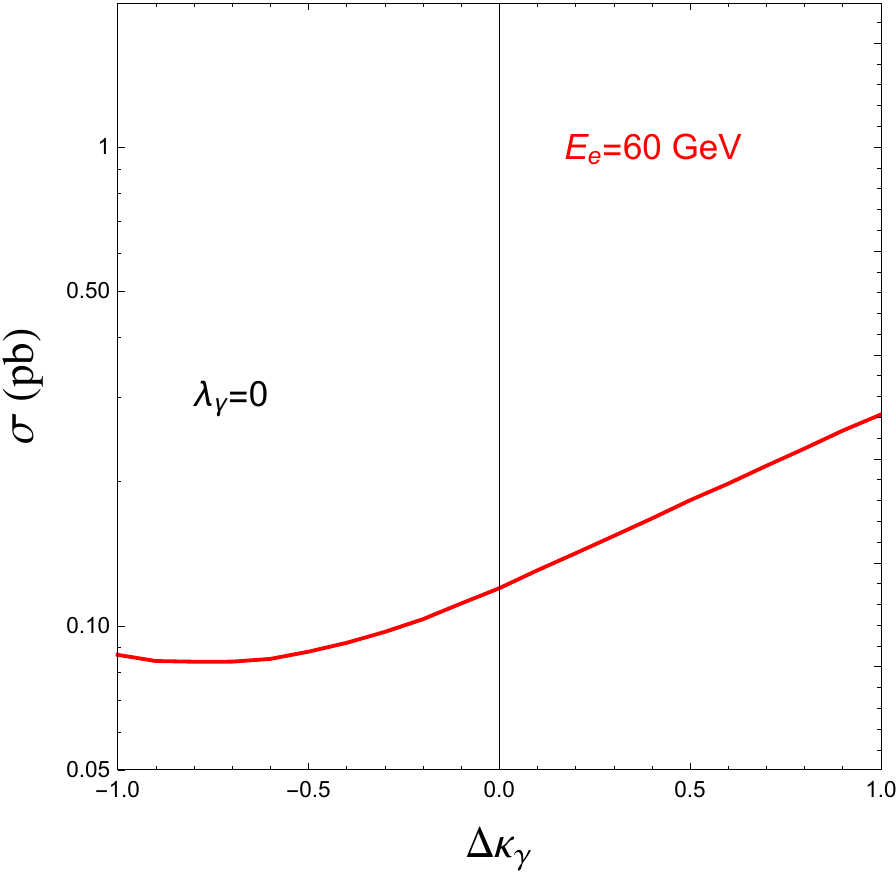}}
\caption{the total cross sections $\sigma_{e^{-}p\rightarrow e^{-}\mu^{+}\nu_{\mu}j}~(\sigma_{tot})$ varying with $\lambda_{\gamma}$~(left panel) and $\Delta\kappa_{\gamma}$~(right panel) }
\label{signals-sigmas-TGCs}
\end{figure}

In Fig.\ref{signals-sigmas-TGCs} we plot the total cross sections $\sigma_{tot}$ of the $e^{-}p\rightarrow e^{-}\mu^{+}\nu_{\mu}j$ process. The basic cuts are
\bea
&&|\eta_{\ell,j}|<5 \nonumber \\ 
&&\Delta R_{\ell\ell}>0.4   \nonumber \\
&&\Delta R_{\ell j}>0.4 \\ 
&&P_{T\ell}>10~\rm{GeV} \nonumber \\
&&P_{Tj}>20~\rm{GeV}, \nonumber
\eea
where $\ell$ and $j$ mean leptons and jets in the final state, respectively. Off-shell $W^{+}$ contribution is also taken into account for the respect of gauge invariance, though the result is actually dominated by the on-shell $W^{+}$ contribution. The production cross section in the SM is $0.120~pb$ while small aTGC contributes only $\mathcal{O}(fb)$. One can see the $\sigma_{tot}$ increases monotonically with $\Delta\kappa_{\gamma}$ and the absolute value of $\lambda_{\gamma}$ within the parameter region allowed by current experiments, but this is not yet enough to probe tiny aTGC contributions.~Therefore, the kinematic differential distributions are to be used as an indirect probe of the anomalous couplings. 
We would demonstrate this idea by studying the $\cos\theta_{\mu W}$ variable in $W$ boson decay and use it for its polarization information.
In addition, the azimuthal angle $\Delta\phi_{ej}$, which was used to measure the $CP$ nature of Higgs couplings~\cite{Plehn:2001nj}, would  be used as well.

\subsection{Kinematic distributions with aTGCs}
\begin{figure}[H]
\centering
\subfloat[$\Delta\kappa_{\gamma}=0$]{\includegraphics[scale=0.3]{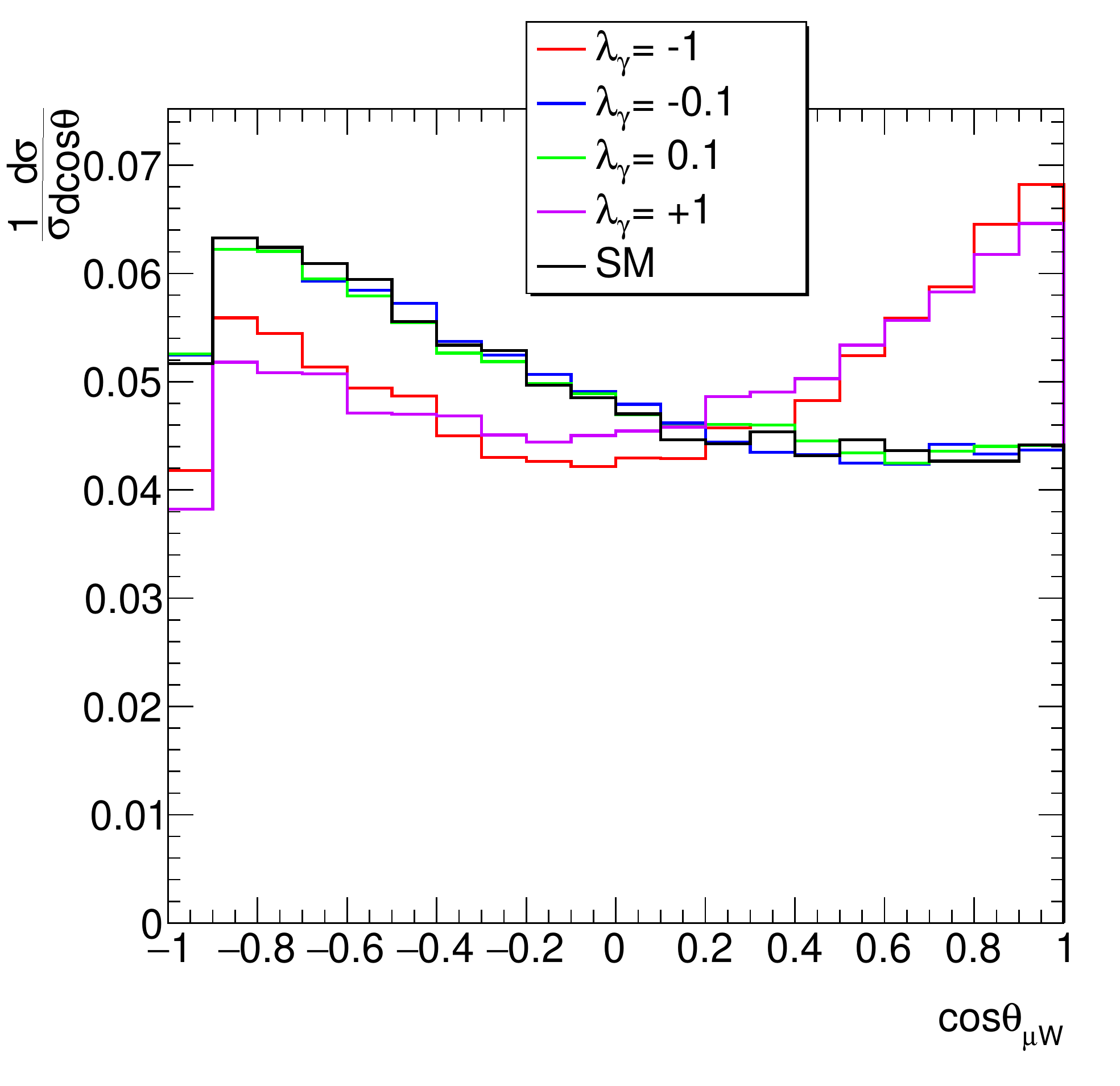}}
\subfloat[$\lambda_{\gamma}=0$]{\includegraphics[scale=0.3]{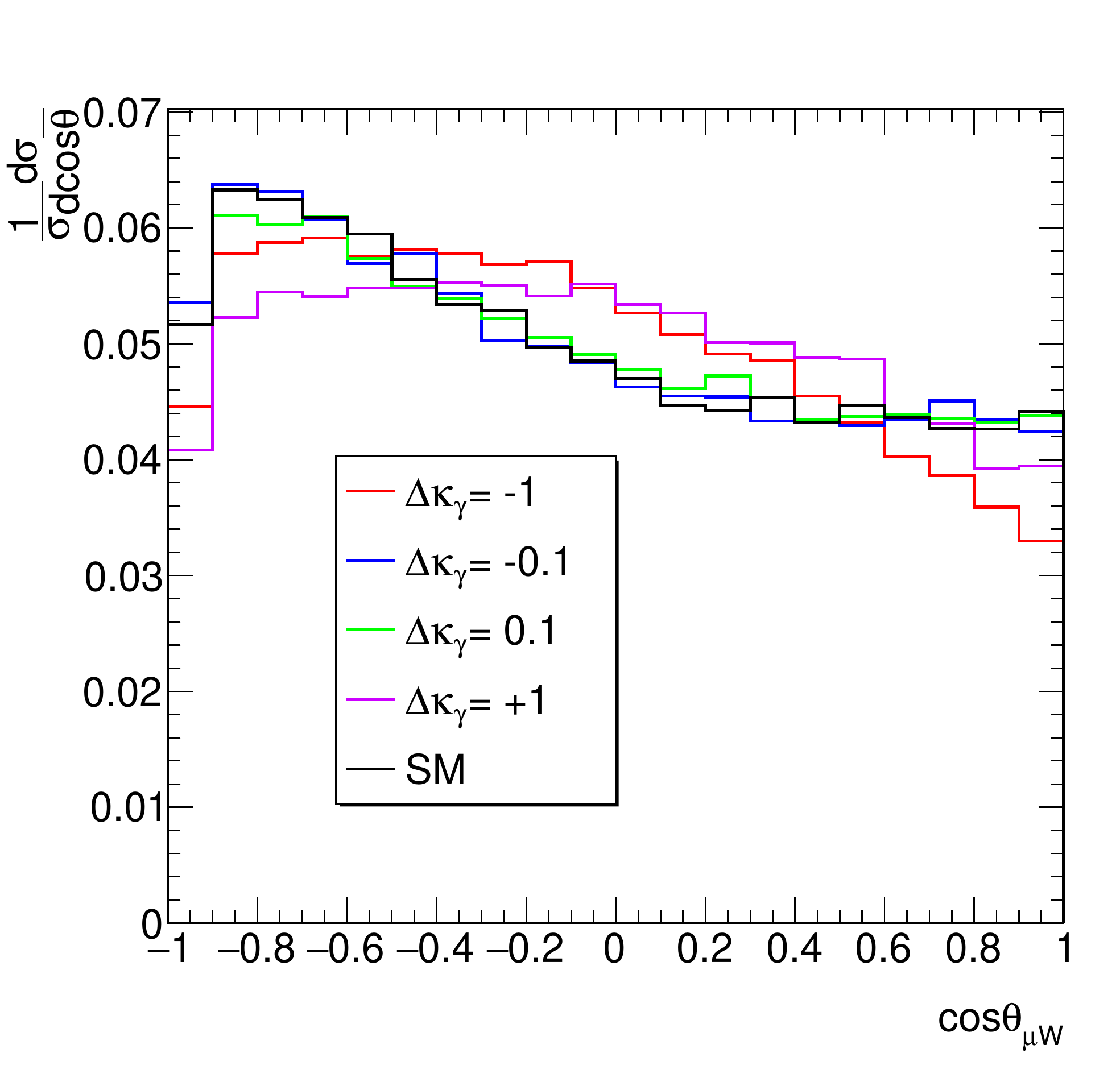}}
\caption{The normalized $\cos\theta_{\mu W}$ distributions varying with $\lambda_{\gamma}$~(left panel) and $\Delta\kappa_{\gamma}$~(right panel) respectively for $E_{e}=60~$GeV.}
\label{cos-lambda-dkappa-60}
\end{figure}

\begin{figure}[H]
\centering
\subfloat[$\Delta\kappa_{\gamma}=0$]{\includegraphics[scale=0.3]{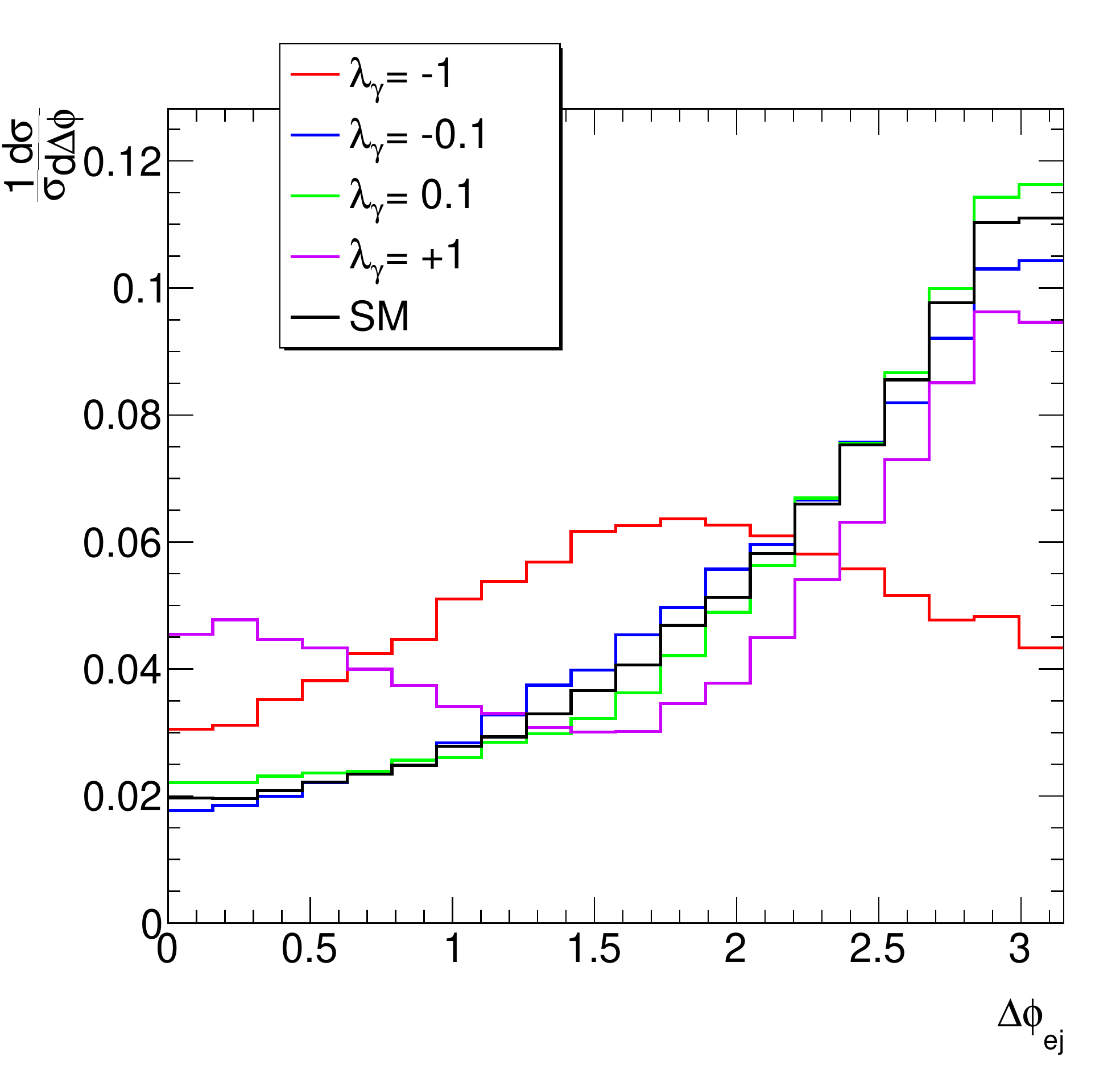}}\qquad
\subfloat[$\lambda_{\gamma}=0$]{\includegraphics[scale=0.3]{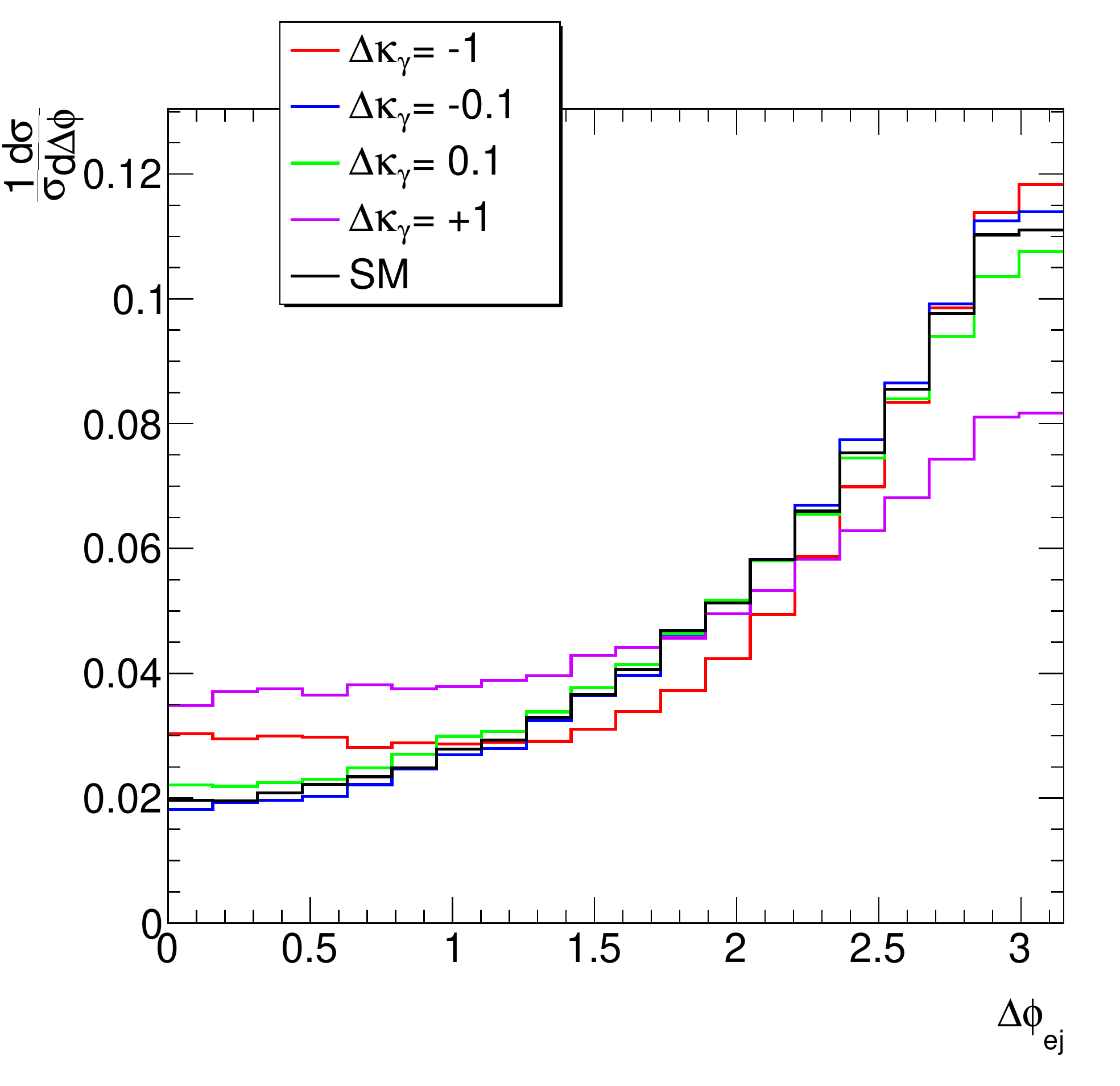}}
\caption{The normalized $\Delta\phi_{ej}$ distributions varying with $\lambda_{\gamma}$~(left panel) and $\Delta\kappa_{\gamma}$~(right panel)~respectively for $E_{e}=60~$GeV.}
\label{dphi-lambda-dkappa-60}
\end{figure}

We turn to more detailed discussion on $\cos\theta_{\mu W}$ and $\Delta\phi_{ej}$ distributions in the $e^{-}p\rightarrow e^{-}\mu^{+}\nu_{\mu}j$ process with nonvanishing $\lambda_{\gamma}$ and $\Delta\kappa_{\gamma}$. For concreteness, $\theta_{\mu W}$ is defined as the angle between the decay product $\mu^+$ in the $W^+$ rest frame and $W^+$ direction in the collision rest frame. $\Delta\phi_{ej}$ is the angle between scattered beam electron and parton on the azimuthal plane. In Fig.\ref{cos-lambda-dkappa-60} and Fig.\ref{dphi-lambda-dkappa-60}, we show $\cos\theta_{\mu W}$ and $\Delta\phi_{ej}$ distributions varying with $\lambda_{\gamma}$ and $\Delta\kappa_{\gamma}$ when $E_{e}=60$~GeV, where the red, blue, green, purple and black lines correspond to the $\lambda_{\gamma}/\Delta\kappa_{\gamma}=-1, -0.1, 0.1, +1$ and 0~(SM) respectively. 

According to the semiquantitive description of the $e^{-}p\rightarrow e^{-}W^{+}j$ process with the helicity technique~\cite{Baur:1989gh}, the 
aTGC $\lambda_{\gamma}$ leads to a significant enhancement in the transverse polarization fraction of the $W$ boson, while $\Delta\kappa_{\gamma}$ leads to a similar enhancement in the longitudinal component fraction. This could be seen from the $\cos\theta_{\mu W}$ distribution in Fig.\ref{cos-lambda-dkappa-60}. The black line shows that $\mu^+$ tends to move in direction opposite of the $W^+$ boson when there's no aTGC contribution. In the left panel, qualitative change, that the peak moves from $\cos\theta_{\mu W}=-1$ to $\cos\theta_{\mu W}=1$ as the aTGC terms dominate, could be seen in the red/purple lines. In the meantime, the peak in the right panel moves to $\cos\theta_{\mu W}\simeq 0$ due to larger contribution from longitudinal-polarized $W$ when $\Delta\kappa_{\gamma}$ is contributing. In both panels, the distributions for $|\lambda_{\gamma}/\Delta\kappa_{\gamma}|=0.1$ are quite similar to the SM distribution, indicating we have to use a more precise method, e.g. the $\chi^2$ method, to measure tiny but nonzero aTGC values.

On the other hand, the $\Delta\phi_{ej}$ distribution would show a peak at $\Delta\phi_{ej}=\pi$ without contribution from aTGCs. That is to say, in the SM the scattered $e^{-}$ and jet are dominantly back-to-back on the azimuthal plane. Just like $\cos\theta_{\mu W}$ , the $\Delta\phi_{ej}$ would present a deviation from the SM in its distribution with $\lambda_{\gamma}$ and $\Delta\kappa_{\gamma}$, as is shown in Fig.\ref{dphi-lambda-dkappa-60}.
We also notice that when $|\lambda_{\gamma}|$ is large~($\lambda_{\gamma}=\pm1$), the shape of the $\Delta\phi_{ej}$ distribution depends on the sign of $\lambda_{\gamma}$: (i) $\lambda_{\gamma}=+1$, the $\Delta\phi_{ej}$ distribution has two peaks at $\Delta\phi_{ej}=0/\pi$ as part of the $e^{-}$ and jet now move in the same direction on the azimuthal plane; (ii) $\lambda_{\gamma}=-1$, the maximum of distribution shifts to around $\Delta\phi_{ej}=\frac{\pi}{2}$.

\subsection{Reconstruction of $W^+$ in $e^{-}p\rightarrow e^{-}W^{+}j$}
At the LHeC, the collision energy is asymmetric, and the final states mostly move toward the proton beam direction. Moreover, in such a $e^-p$ collision, the momentum conservation condition in the $z$ direction cannot be used as a result of the unknown Bjorken $x$. Therefore, reconstructing full final states with the $W$ boson-invariant mass and the massless neutrino is quite difficult because there're always two solutions for the invisible neutrino. One way to distinguish them is to assume $W$ decay products would move in the same direction and have a small angular difference. Then the solution with momentum more parallel with the muon is used to reconstruct the $W$ boson.

In addition, we could also get a single accurate solution for the invisible neutrino by combining energy and $z$-direction momentum conservation conditions to cancel unknown Bjorken $x$ dependence. Splitting the final states into two parts, the invisible neutrino with $p^{\mu}_{\nu_{\mu}}$ and the others($e^{-}$, $\mu^{+}$ and jet) with $p^{\mu}_{e'j\mu}$, after a bit more algebra which is shown in the Appendix A, we have
 \bea
\centering
p_{\nu_{\mu}}^{z}=\frac{(2E_{e}-E_{e'j\mu}-p_{e'j\mu}^{z})^{2}-(p_{\nu_{\mu}}^{T})^{2}}{2(2E_{e}-E_{e'j\mu}-p_{e'j\mu}^{z})},
\label{neutrino mom}
\eea
where $p_{\nu_{\mu}}^{T}$ is the transverse momentum of the neutrino i.e the missing transverse energy $/\kern-0.57em E_{T}$, $E_{e}$ is the energy of the initial electron. This avoids the ambiguity of two solutions.

\begin{figure}[H]
\centering
\includegraphics[scale=0.4]{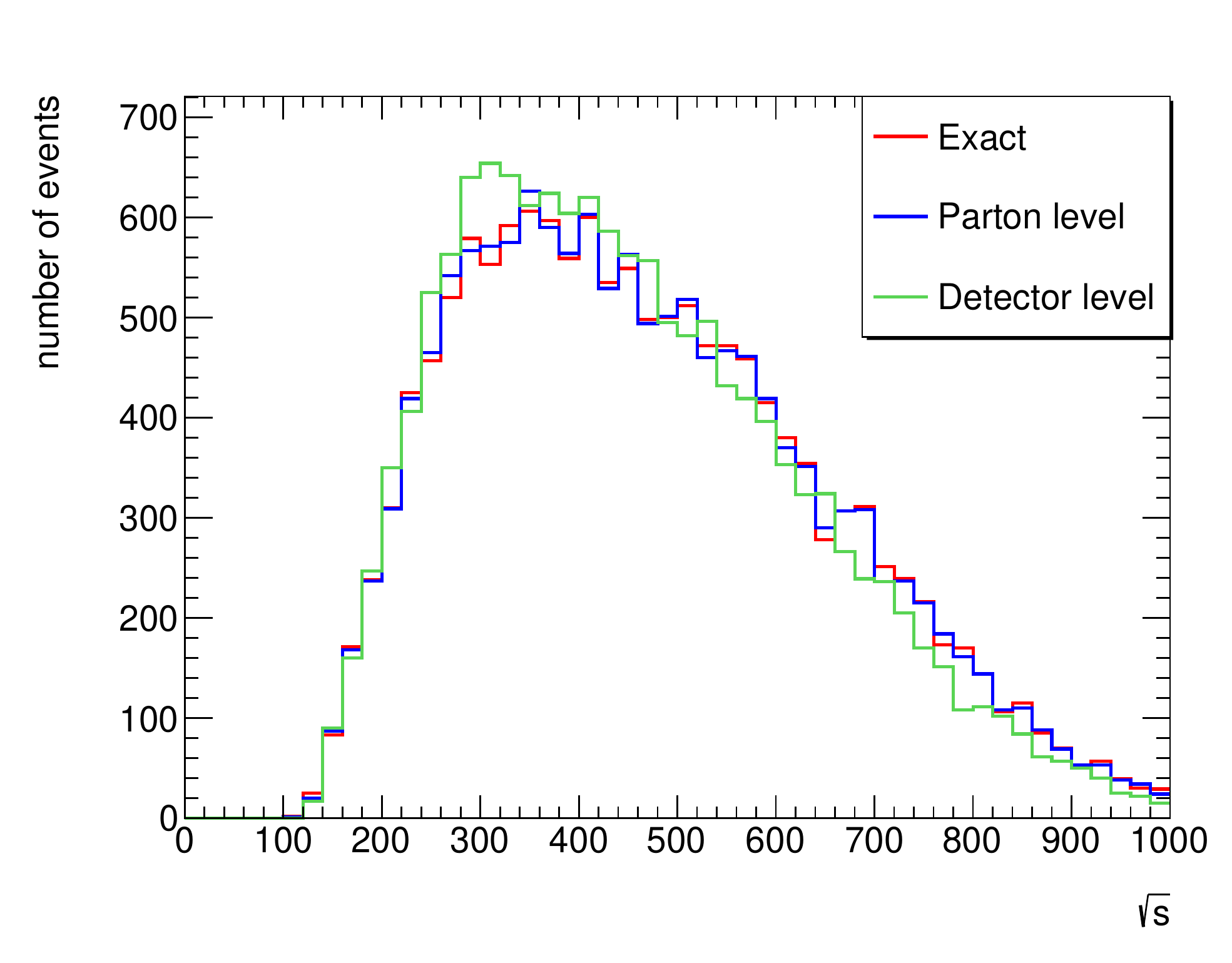}
\caption{Comparison of partonic collision energy $\sqrt{\hat{s}}$ distributions of exact value(red), parton-level value(blue) and detector-level value(green). The parton shower and hadronization are simulated with {\it Pythia 6.420}~\cite{Sjostrand:2006za}. The detector simulation is with {\it Delphes 3.3.0}~\cite{deFavereau:2013fsa}.  }
\label{sqrts_del}
\end{figure}

Another kinematic method is the recoil mass, which was used in the Higgs-strahlung process at the $e^{+}e^{-}$ collider~\cite{Juste:1999xv}.~The final states could be separated into two parts: a scattered electron-jet system with $p^{\mu}_{e'j}$ and all remaining particles with $p^{\mu}_X$ called the recoil system. Then, we have 
\bea
\centering
M_{X}^{2}=\hat{s}+M_{e'j}^{2}-2E_{e'j}(E_{q}+E_{e})+2p_{e'j}^{z}(E_{e}-E_{q}),
\label{recoil-mass}
\eea
where $M_{X}$ is the recoil mass, $\hat{s}$ is the partonic collision energy square, and $E_{q}$ and $E_{e}$ are the energy of initial parton and electron.
Since the process we study gets a large contribution from on-shell $W$ channels, we could simply choose the $W$ boson itself as the recoil system and get a relation of Bjorken $x$ with the known input
\bea
x=\frac{M_{W}^{2}-M_{e'j}^{2}+2E_{e}(E_{e'j}-p_{e'j}^{z})}{2E_{P}(2E_{e}-E_{e'j}-p_{e'j}^{z})}.
\eea
With this relation, one could solve for the invisible neutrino because $z$-direction momentum conservation condition is now available.~The explicit procedure is shown in the Appendix A. This method works well for events with an on-shell $W$, but leads to certain deviation for other backgrounds.~{By the way, the above analysis are based on the definition that the z direction is the electron beam moving direction.}~The reconstructed partonic collision energy distributions are shown in Fig.\ref{sqrts_del} through the above relation to confirm the validity of the recoil mass method.

The VBF final state consists of only one forward energetic quark. However, the additional jets due to gluon radiation are still inevitable although most of them are soft or colinear to the final-state quark. Therefore, one would need criteria for correct forward jet tagging, for instance, with jet energy or jet rapidity, etc. In this study, for simplicity, we only select the events with one forward jet and veto all the others with a second or more hard jets ($P_{Tj}> 20~$GeV) to minimize the mistag rate of jets. Under these criteria,  a full simulation including {\it Pythia 6.420} and {\it Delphes 3.3.0} approximately results in a 30\% survival probability. 

\section{results}
\label{result}
Without real data, it is always difficult to do a comprehensive analysis of uncertainties. For instance, one of the leading theoretical uncertainties of SM prediction is the PDF variation. We estimate this contribution in the cross section measurement is 0.6\% with {\it NNPDF23\underline{\hspace{0.5em}}nlo\underline{\hspace{0.5em}}as\underline{\hspace{0.5em}}0119}~sets. On the other hand, one of the purposes of the LHeC is to provide precision measurements of valence quark distributions.~The striking improvement of PDF determinations would lead to a dramatic reduction in the above uncertainty by a factor of three to four in the $\mathcal{O}(10^{-2})$ $x$ region of our processes~\cite{AbelleiraFernandez:2012cc}.~Therefore, aTGC contributions would not be submerged by the PDF uncertainty and one could combine them for the constraints.~We expect this has only an insignificant effect on aTGC constraints and therefore neglect the PDF uncertainty in the following study.

{In the meantime, we set $\slashed{E}_{T}>20~\rm{GeV}$ to avoid pileup errors because of the low transverse energy basic cut before constraining the aTGC bounds. This additional cut results in about 87\% survival probability.~Since the lepton/neutrino $p_{T}$ depends on the polarization of the $W$ boson, the $\slashed{E}_{T}$ cut certainly affects the $\cos\theta_{\mu W}$ distribution. Those events with a neutrino moving in the direction opposite of the $W$ boost direction are likely to be cut away by this cut which corresponds to the $\cos\theta_{\mu W}$ toward 1. We expect it to give a minor improvement on the results.   }

To illustrate the feature of the two kinematic distributions proposed above,  we adopt the $\chi^{2}$ method for large event numbers by assuming that the best-fitting aTGC values of future data equal zero~\cite{Kuss:1996te}. 
\bea
\chi^{2}\equiv\sum_{i}\left(\frac{N_{i}^{BSM}-N_{i}^{SM}}{\sqrt{N_{i}^{SM}}}\right)^{2},
\eea
where $N_{i}^{BSM}$ and $N_{i}^{SM}$ are the numbers of events in the $i$th bin for the differential distributions with and without aTGCs. In this $\chi^2$ method, we use ten bins to analysis the distributions and take 95\%~C.L. bounds as the aTGC values. Single-parameter fitting results at parton level are shown in Table~\ref{limit-tgcs} with two electron beam energy options and $\mathcal{L}=1~ab^{-1}$ integrated luminosity. Two aTGC parameter bounds are pushed to a few $\mathcal{O}(10^{-3})$ level in the most ideal case when there's an upgrade for $E_{e}=140~$GeV. In the best measurement channel, we find that $\Delta\phi_{ej}$ would impose stringent constraints on both $\lambda_{\gamma}$ and $\Delta\kappa_{\gamma}$. The other observable $\cos\theta_{\mu W}$, however, could put a tight bound on $\Delta\kappa_{\gamma}$ but fails to constrain $\lambda_{\gamma}$. Moreover, the $\mu^{+}$ channel is indeed more sensitive to aTGCs than the $\mu^{-}$ channel as we have discussed in section~\ref{pheno}.

\begin{figure}[H]
\centering
\includegraphics[scale=0.4]{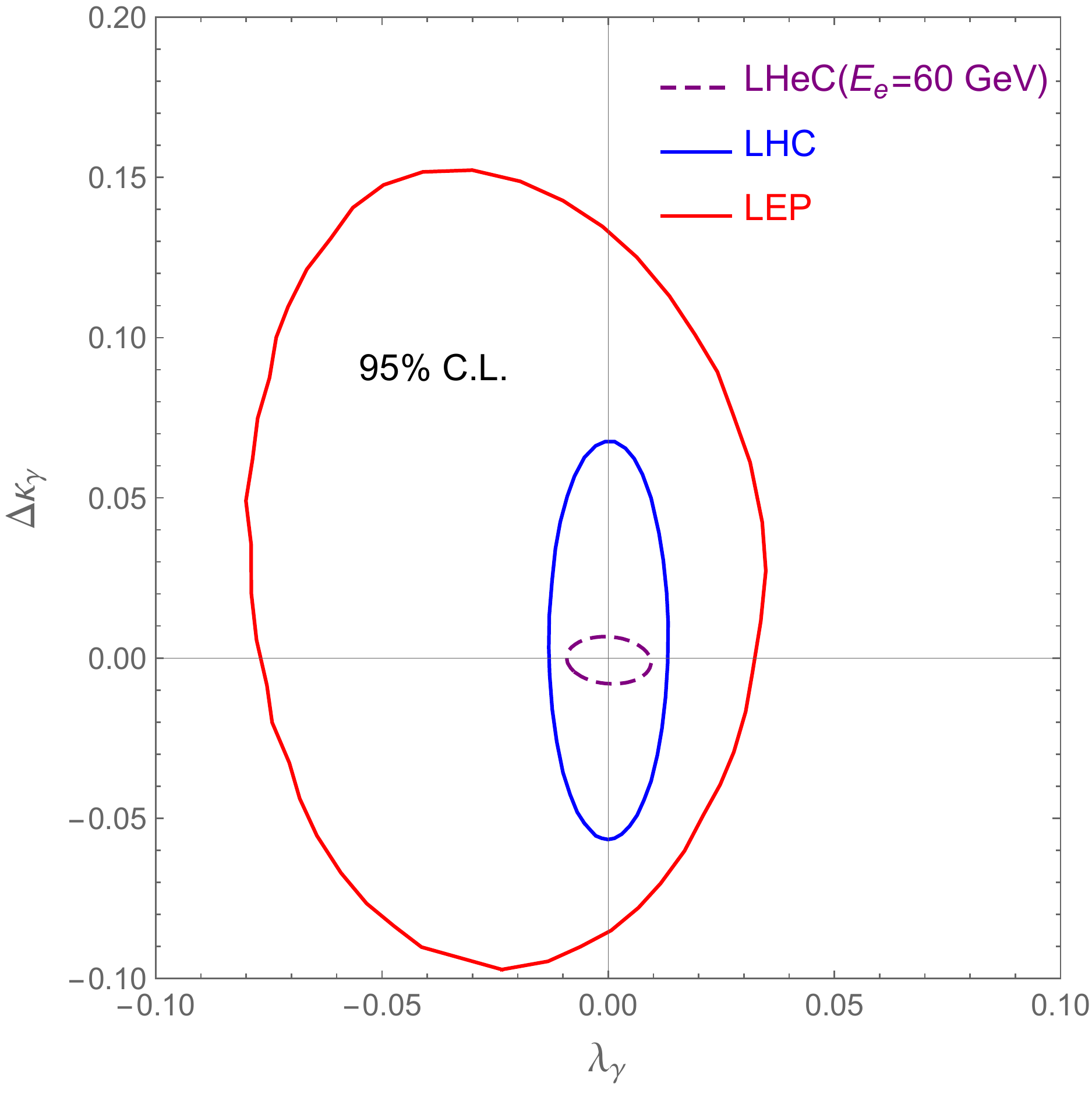}
\caption{Two-parameter fitting results of aTGC bounds at 95\% C.L. for the LHeC, LHC, and LEP.}
\label{bounds-contour}
\end{figure}

\begin{table}[H]
\begin{center}
\begin{tabular}{|c|c|c|c|c|c|}
\hline
\multicolumn{1}{|c|}{\multirow{2}{*}{\diagbox{parameter}{variable}} }&\multicolumn{2}{|c|}{$\mu^{+}$ decay, $E_{e}=60~$GeV}&\multicolumn{2}{|c|}{$\mu^{+}$ decay, $E_{e}=140~$GeV}&\\\cline{2-6}
& $\cos\theta_{\mu^+ W^+}$ & $\Delta\phi_{ej}$ &$\cos\theta_{\mu^+ W^+}$ &$\Delta\phi_{ej}$ &SM \\\hline
$\lambda_{\gamma}$ & $\times$ & [-0.007, 0.0056] & $\times$ &[-0.0034, 0.0021]& 0\\\hline
$\Delta\kappa_{\gamma}$  & [-0.0054, 0.006] & [-0.0043, 0.0054] &[-0.002, 0.0017]  &[-0.003, 0.0021]&0 \\\hline\hline
\multicolumn{1}{|c|}{\multirow{2}{*}{\diagbox{parameter}{variable}} }&\multicolumn{2}{|c|}{$\mu^{-}$ decay, $E_{e}=60~$GeV}&\multicolumn{2}{|c|}{$\mu^{-}$ decay, $E_{e}=140~$GeV}&\\\cline{2-6}
& $\cos\theta_{\mu^- W^-}$ & $\Delta\phi_{ej}$ &$\cos\theta_{\mu^- W^-}$ &$\Delta\phi_{ej}$ &SM \\\hline
$\lambda_{\gamma}$ & $\times$ & [-0.0092, 0.0096] & $\times$ &[-0.0031, 0.0045]& 0\\\hline
$\Delta\kappa_{\gamma}$  & [-0.0073, 0.0071] &  [-0.0067, 0.0075]&[-0.0016, 0.0024]  &[-0.004, 0.0043]&0 \\\hline
\end{tabular}
\caption{The 95\%~C.L. bound on aTGC $\lambda_{\gamma}$ and $\Delta\kappa_{\gamma}$, obtained from the kinematic observables $\cos\theta_{\mu^{\pm} W^{\pm}}$ and $\Delta\phi_{ej}$ at the LHeC with $E_{e}=60~$ and 140~GeV. The results listed are from single-parameter fitting when the other one is fixed to its SM value. The  ``$\times$'' in the table means this bound is no better than the ones from the LEP.}
\label{limit-tgcs}
\end{center}
\end{table}
In Fig.\ref{bounds-contour}, we show the two-parameter $\Delta\phi_{ej}$ fitting result with default $E_e=$~60~GeV~(purple dashed line) on the $\lambda_{\gamma}$--$\Delta\kappa_{\gamma}$ plane.
For comparison, we also include the present LHC~(blue solid line) and LEP~(red solid line) exclusion contours. LHeC result would surpass both existing limits.~What is more, there's also a significant improvement in constraining $\Delta\kappa_{\gamma}$ parameter because the observables we choose are sensitive to the enhancement in longitudinal polarization.

The above results are all obtained via pure partonic level study which is certainly unrealistic. However, as we have discussed 
in the previous section, the criteria of vetoing a second or more hard jets minimizes the mistag rate and only gives about 30\% survival probability. Therefore, to achieve the same results in a full simulation with ${\it Pythia}$ and ${\it Delphes}$, one expects 
about threefold integrated luminosity.

\section{conclusion}
We find in the $e^{-}p\rightarrow e^{-}\mu^{+}\nu_{\mu}j$ subchannel, the sensitivity to $\lambda_{\gamma}$ and $\Delta\kappa_\gamma$ could reach $\mathcal{O}(10^{-3}$) when $\mathcal{L}=1~ab^{-1}$ based on the $\chi^{2}$ method at parton level with the expectation of more precise PDFs at the future LHeC, while in a full simulation the integrated luminosity needs to be increased to 2-3$~ab^{-1}$ to be consistent with the result. Furthermore, the same result might be reached with approximately half integrated luminosity if we combined the $\mu^{+}$ and $\mu^{-}$ channels. From the results in Table~\ref{limit-tgcs} and Fig.\ref{bounds-contour}, we could see a significant improvement compared to the present LHC and LEP bounds. Therefore, the measurement of the $e^{-}p\rightarrow e^{-}W^{\pm}j$ process at the LHeC would provide a promising opportunity to probe aTGCs and improve our knowledge of the gauge sector. For future aTGC measurement, we expect complementary studies with different electron beam polarizations and more realistic detector-level analysis to be helpful. 

With regard to more technical analysis methods, we may further consider the joint distribution of $\Delta\phi_{ej}$ and $W$ boson polorization, which could be realized by dividing each $\Delta\phi_{ej}$ bin into three sub-bins corresponding to three $W$ boson polarization states with fractions $f_L, f_R, \text{and} f_0$ respectively~\cite{Bern:2011ie}. On the other side, these polarization fractions are also able to be calculated by decomposing the $\cos\theta_{\mu W}$ distribution in Legendre polynomails of $\cos\theta_{\mu W}$. 
 
Finally, it is noteworthy that the kinematic methods in event reconstruction, through which one could retrieve z-direction momentum conservation condition despite of the ignorance of initial state parton and final state neutrino momentums. We believe the kinematic methods are useful for future measurements of processes with $\slashed{E}_T$ at this $ep$ collider. 

\section{Acknowledgements}
We thank Ligong Bian, Ran Ding, Ke-pan Xie, Bin Yan and Kechen Wang for helpful discussion. And the work is supported in part by the National Science Foundation of China~(11422544, 11535002, 11375151, 11535002)~and the Zhejiang University Fundamental Research Funds for the Central Universities. KW is also supported by Zhejiang University K.P Chao High Technology Development Foundation.

\appendix
  \renewcommand{\appendixname}{Appendix}

  \section{Neutrino momentum reconstruction}
 \begin{figure}[H]
\centering
{\includegraphics[scale=0.7]{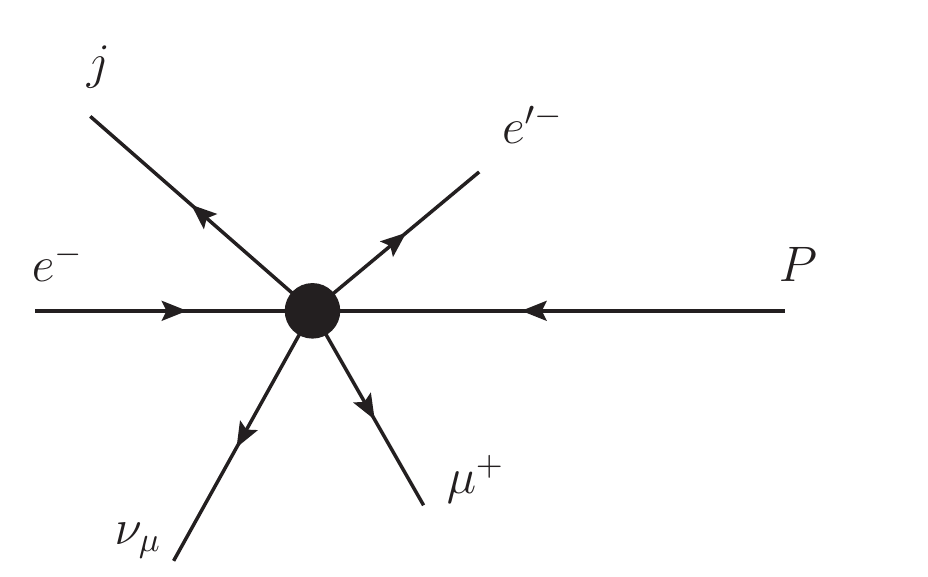}}\qquad
\caption{The collision process at LHeC.}
\label{collision}
\end{figure}
In the initial state,
\bea
&&p^{\mu}_{e}\nonumber: (E_{e},0,0,p^{z}_{e})\\ \nonumber
&&p^{\mu}_{P}: (E_{P},0,0,-p^{z}_{P}), \nonumber
\eea
where $E_{e}=p^{z}_{e}$ and $E_{P}=-p^{z}_{P}$;

In the final state,
\bea
&&p^{\mu}_{e}\nonumber:(E_{e'},p_{e'}^{x},p_{e'}^{y},p_{e'}^{z})\\ \nonumber
&&p^{\mu}_{j}:(E_{j},p_{j}^{x},p_{j}^{y},p_{j}^{z}) \\ \nonumber
&&p^{\mu}_{\mu^{+}}:(E_{\mu},p_{\mu}^{x},p_{\mu}^{y},p_{\mu}^{z}) \\ \nonumber
&&p^{\mu}_{\nu_{\mu}}:(E_{\nu_{\mu}},p_{\nu_{\mu}}^{x},p_{\nu_{\mu}}^{y},p_{\nu_{\mu}}^{z}). \nonumber
\eea

\subsection{Method 1: Energy-momentum conservation}
We split the final states into two parts: the invisible neutrino with $p^{\mu}_{\nu_{\mu}}$ and the others($e^{-}$, $\mu^{+}$, and jet) with $p^{\mu}_{e'j\mu}$. Then, we have 
\bea
p_{\nu_{\mu}}^{T}&=&\sqrt{(p_{\nu_{\mu}}^{x})^{2}+(p_{\nu_{\mu}}^{y})^{2}}=\slashed{E}_{T}; \nonumber \\
E_{\nu_{\mu}}&=&\sqrt{(p_{\nu_{\mu}}^{T})^{2}+(p_{\nu_{\mu}}^{z})^{2}};  \\
E_{e'\mu j}&=&E_{e'}+E_{\mu}+E_{j}; \nonumber \\
p_{e'\mu j}^{z}&=&p_{e'}^{z}+p_{\mu}^{z}+p_{j}^{z}. \nonumber 
\eea
Using the energy-momentum conservation between the initial and final states, we get
 \bea
&&E_{e}+E_{P}=E_{e'}+E_{\mu}+E_{\nu_{\mu}}+E_{j}=E_{e'\mu j}+\sqrt{(p_{\nu_{\mu}}^{T })^{2}+(p_{\nu_{\mu}}^{z})^{ 2}};\\
&&p_{e}+p_{P}=E_{e}-E_{P}=p_{e'}^{z}+p_{\mu}^{z}+p_{\nu_{\mu}}^{z}+p_{j}^{z}=p_{e'\mu j}^{z}+p_{\nu_{\mu}}^{z}.
\eea

After the combination of Eqs. (A1) and (A2) with a little algebra, we can get a single accurate solution of the z direction momentum of the invisible neutrino:
\bea
(\text{A2})+(\text{A3}) \nonumber
&&\Rightarrow
2E_{e}=E_{e'\mu j}+p_{e'\mu j}^{z}+\sqrt{(p_{\nu_{\mu}}^{T })^{2}+(p_{\nu_{\mu}}^{z})^{ 2}}+p_{\nu_{\mu}}^{z}\\ \nonumber
&&\Rightarrow
2E_{e}-E_{e'\mu j}-p_{e'\mu j}^{z}-p_{\nu_{\mu}}^{z}=\sqrt{(p_{\nu_{\mu}}^{T })^{2}+(p_{\nu_{\mu}}^{z})^{ 2}}\\ \nonumber
&&\Rightarrow
(2E_{e}-E_{e'\mu j}-p_{e'\mu j}^{z})^{2}+(p_{\nu_{\mu}}^{z})^{2}-2 p_{\nu_{\mu}}^{z} (2E_{e}-E_{e'\mu j}-p_{e'\mu j}^{z})=(p_{\nu_{\mu}}^{T })^{2}+(p_{\nu_{\mu}}^{z})^{ 2}\\ 
&&\Rightarrow
p_{\nu_{\mu}}^{z}=\frac{(2E_{e}-E_{e'j\mu}-p_{e'j\mu}^{z})^{2}-(p_{\nu_{\mu}}^{T})^{2}}{2(2E_{e}-E_{e'j\mu}-p_{e'j\mu}^{z})}, (2E_{e}-E_{e'j\mu}-p_{e'j\mu}^{z} \neq 0) 
\eea

\subsection{Method 2: Recoil mass}
First, the final states could be separated into two parts: a scattered electron-jet system with $p^{\mu}_{e'j}$ and all remaining particles with $p^{\mu}_X$ called the recoil system.~Since the process we study gets a large contribution from on-shell the $W$ channels, we could simply choose the $W$ boson itself as the recoil system. In the partonic level, we have 
\bea
&&p^{\mu}_{q}\equiv xp^{\mu}_{P}:(E_{q},0,0,p_{q}^{z}) \nonumber \\
&&p^{\mu}_{e'j}\equiv p^{\mu}_{e'}+p^{\mu}_{j}:(E_{e'j},p_{e'j}^{x},p_{e'j}^{y},p_{e'j}^{z})   \\
&&p^{\mu}_{X} \equiv p^{\mu}_{q}+p^{\mu}_{e}-p^{\mu}_{e'j}:(E_{X},-p_{e'j}^{x},-p_{e'j}^{x},p_{X}^{z}), \nonumber
\eea
where $q$ is the parton from the initial proton and $x$ is the unknown Bjorken parameter. Then we can calculate the partonic collision energy square $\hat{s}$ of this process,
\bea
\hat{s}=(p_{e}^{\mu}+p_{q}^{\mu})^{2}=4E_{e}E_{q}=4xE_{e}E_{P},
\eea
as well as compute $\hat{s}$ through the final-state particles:
\bea
\hat{s}&=&(p_{X}^{\mu}+p_{e'j}^{\mu})^{2} \nonumber \\
&=&M_{X}^{2}+M_{e'j}^{2}+2\left[E_{x}E_{e'j}+(p_{e'j}^{x})^{2}+(p_{e'j}^{y})^{2}-p_{X}^{z}p_{e'j}^{z}\right] \nonumber \\
&=&M_{X}^{2}+M_{e'j}^{2}+2\left[E_{e'j}(E_{q}+E_{e}-E_{e'j})-p_{e'j}^{z}(p_{q}^{z}+p_{e}^{z}-p_{e'j}^{z})+(p_{e'j}^{x})^{2}+(p_{e'j}^{y})^{2}\right] \nonumber \\
&=&M_{X}^{2}-M_{e'j}^{2}+2E_{e'j}(xE_{P}+E_{e})-2p_{e'j}^{z}(xp_{P}^{z}+p_{e}^{z})
\eea
where $M_{X} \text{and} M_{e'j}$ are invariant masses of the recoil system and scattered electron-jet system respectively i.e. $M_{X}^{2}=p_{X}^{\mu}\cdot p_{X \mu},M_{e'j}=p_{e'j}^{\mu}\cdot p_{e'j \mu}$. $E_{X}=E_{q}+E_{e}-E_{e'j}$, and $p_{X}^{z}=p_{q}^{z}+p_{e}^{z}-p_{e'j}^{z}$ have been used in the above derivation process. Finally, combining Eqs. (A5) and (A6), and substituting the $W$ boson mass $M_{W}$ for the recoil mass $M_{X}$, we can get the unknown Bjorken $x$:
\bea
(\text{A6})=(\text{A7}) \nonumber
&&\Rightarrow
M_{W}^{2}=4xE_{e}E_{P}+M_{e'j}^{2}-2E_{e'j}(xE_{P}+E_{e})+2p_{e'j^{z}}(xp_{P}^{z}+p_{e}^{z})\\ 
&&\Rightarrow
x=\frac{M_{W}^{2}-M_{e'j}^{2}+2E_{e}(E_{e'j}-p_{e'j}^{z})}{2E_{P}(2E_{e}-E_{e'j}-p_{e'j}^{z})}, (2E_{e}-E_{e'j\mu}-p_{e'j\mu}^{z} \neq 0)
\eea
So, it is easy to get the z direction momentum of the invisible neutrino:
\bea
p_{\nu_{\mu}}^{z}=E_{e}-xE_{P}-p_{e'}^{z}-p_{j}^{z}.
\eea

\begin{thebibliography}{99} 




\bibitem{Aad:2012tfa} 
  G.~Aad {\it et al.} [ATLAS Collaboration],
  Phys.\ Lett.\ B {\bf 716}, 1 (2012)
  doi:10.1016/j.physletb.2012.08.020
  [arXiv:1207.7214 [hep-ex]].
  
  
  
\bibitem{Chatrchyan:2012xdj} 
  S.~Chatrchyan {\it et al.} [CMS Collaboration],
  Phys.\ Lett.\ B {\bf 716}, 30 (2012)
  doi:10.1016/j.physletb.2012.08.021
  [arXiv:1207.7235 [hep-ex]].
 
  

\bibitem{AbelleiraFernandez:2012cc} 
  J.~L.~Abelleira Fernandez {\it et al.} [LHeC Study Group],
  J.\ Phys.\ G {\bf 39}, 075001 (2012)
  doi:10.1088/0954-3899/39/7/075001
  [arXiv:1206.2913 [physics.acc-ph]].


\bibitem{Han:2009pe} 
  T.~Han and B.~Mellado,
  Phys.\ Rev.\ D {\bf 82}, 016009 (2010)
  doi:10.1103/PhysRevD.82.016009
  [arXiv:0909.2460 [hep-ph]].
  

\bibitem{Biswal:2014oaa} 
 S.~S.~Biswal, M.~Patra and S.~Raychaudhuri,
arXiv:1405.6056 [hep-ph].

\bibitem{cakir}
I.T. Cakir, A.~Senol, O.~Cakir, et al. Int. J. Chem. Mol. Nucl. Mater. Metal. Eng., 2016, 9(CERN-ACC-2016-0107): 34-38.
 
 \bibitem{cakir1} 
 I.T. Cakir, {\it et al.},
 Acta. Phys. Pol. B, vol. 45, 2014,
pp. 1947-1962. 

  \bibitem{Baur:1989gh} 
  U.~Baur and D.~Zeppenfeld,
  Nucl.\ Phys.\ B {\bf 325}, 253 (1989).
  doi:10.1016/0550-3213(89)90457-4
  
  \bibitem{Bian:2015zha} 
  L.~Bian, J.~Shu and Y.~Zhang,
  JHEP {\bf 1509}, 206 (2015)
  doi:10.1007/JHEP09(2015)206
  [arXiv:1507.02238 [hep-ph]].

  \bibitem{Hagiwara:1993ck} 
  K.~Hagiwara, S.~Ishihara, R.~Szalapski and D.~Zeppenfeld,
  Phys.\ Rev.\ D {\bf 48}, 2182 (1993).
  doi:10.1103/PhysRevD.48.2182
  \bibitem{Hagiwara:1992eh} 
  K.~Hagiwara, S.~Ishihara, R.~Szalapski and D.~Zeppenfeld,
  Phys.\ Lett.\ B {\bf 283}, 353 (1992).
  doi:10.1016/0370-2693(92)90031-X
  \bibitem{DeRujula:1991ufe} 
  A.~De Rujula, M.~B.~Gavela, P.~Hernandez and E.~Masso,
  Nucl.\ Phys.\ B {\bf 384}, 3 (1992).
  doi:10.1016/0550-3213(92)90460-S
  
  \bibitem{Baur:1987mt} 
  U.~Baur and D.~Zeppenfeld,
  Phys.\ Lett.\ B {\bf 201}, 383 (1988).
  doi:10.1016/0370-2693(88)91160-4  

\bibitem{Corbett:2017qgl} 
  T.~Corbett, O.~J.~P.~Eboli and M.~C.~Gonzalez-Garcia,
  Phys.\ Rev.\ D {\bf 96}, no. 3, 035006 (2017)
  doi:10.1103/PhysRevD.96.035006
  [arXiv:1705.09294 [hep-ph]].
  
   
    \bibitem{Schael:2013ita} 
  S.~Schael {\it et al.} [ALEPH and DELPHI and L3 and OPAL and LEP Electroweak Collaborations],
  Phys.\ Rept.\  {\bf 532}, 119 (2013)
  doi:10.1016/j.physrep.2013.07.004
  [arXiv:1302.3415 [hep-ex]].

    \bibitem{cms-2017-tgcs} 
  A.~M.~Sirunyan {\it et al.} [CMS Collaboration],
  Phys.\ Lett.\ B {\bf 772}, 21 (2017)
  doi:10.1016/j.physletb.2017.06.009
  [arXiv:1703.06095 [hep-ex]]. 

\bibitem{Aaboud:2017cgf} 
  M.~Aaboud {\it et al.} [ATLAS Collaboration],
  arXiv:1706.01702 [hep-ex].

  
\bibitem{Alwall:2014hca} 
  J.~Alwall {\it et al.},
  JHEP {\bf 1407}, 079 (2014)
  doi:10.1007/JHEP07(2014)079
  [arXiv:1405.0301 [hep-ph]].

  
   \bibitem{Plehn:2001nj} 
  T.~Plehn, D.~L.~Rainwater and D.~Zeppenfeld,
 Phys.\ Rev.\ Lett.\  {\bf 88}, 051801 (2002)
  doi:10.1103/PhysRevLett.88.051801
  [hep-ph/0105325].
  
  
  
\bibitem{Juste:1999xv} 
  A.~Juste,
  hep-ex/9912041.

\bibitem{Sjostrand:2006za} 
  T.~Sjostrand, S.~Mrenna and P.~Z.~Skands,
  JHEP {\bf 0605}, 026 (2006)
  doi:10.1088/1126-6708/2006/05/026
  [hep-ph/0603175].

\bibitem{deFavereau:2013fsa} 
  J.~de Favereau {\it et al.} [DELPHES 3 Collaboration],
  JHEP {\bf 1402}, 057 (2014)
  doi:10.1007/JHEP02(2014)057
  [arXiv:1307.6346 [hep-ex]].


\bibitem{Kuss:1996te} 
  I.~Kuss and D.~Schildknecht,
  Phys.\ Lett.\ B {\bf 383}, 470 (1996)
  doi:10.1016/0370-2693(96)00770-8
  [hep-ph/9603283].

\bibitem{Bern:2011ie} 
  Z.~Bern, G.~Diana {\it et al.},
  Phys.\ Rev.\ D {\bf 84}, 034008 (2011)
  doi:10.1103/PhysRevD.84.034008
  [arXiv:1103.5445 [hep-ph]].


  
  \end{thebibliography}
\end{document}